\documentstyle[prb,aps,amsfonts,amssymb,multicol,epsfig]{revtex}
\renewcommand{\narrowtext}{\noindent\begin{multicols}{2}\noindent
\global\columnwidth20.5pc}
\renewcommand{\widetext}{\end{multicols}
\global\columnwidth42.5pc}  
\renewcommand{\top}[1]{%
 \vskip #1%
 \begin{picture}(290,80)(80,500)%
 \thinlines%
 \put(65,500){\line( 1, 0){255}}\put(320,500){\line( 0, 1){5}}%
 \end{picture}%
}
\newcommand{\bottom}[1]{%
 \vskip #1%
 \begin{picture}(290,80)(80,500)%
 \thinlines%
 \put(330,500){\line( 1, 0){255}}\put(330,500){\line( 0, -1){5}}%
 \end{picture}%
}

\begin{document}

{
\title{Anderson impurity model at finite Coulomb interaction U:
generalized Non-crossing Approximation}
\author{K. Haule$^{1,2}$, S. Kirchner$^2$, J. Kroha$^2$ and P. W\"olfle$^2$}
\address{$^1$J. Stefan Institute, SI-1000 Ljubljana, Slovenia\\ 
$^2$Institut f\"{u}r Theorie der
Kondensierten Materie, Universit\"{a}t Karlsruhe, D-76128
Karlsruhe, Germany }
\date{\today}
\maketitle

\begin{abstract}
We present an extension of the non-crossing approximation (NCA), which
is widely used to calculate properties of Anderson impurity models in
the limit of infinite Coulomb repulsion $U\rightarrow\infty$, to the
case of finite $U$.  A self-consistent conserving pseudo--particle
representation is derived by symmetrizing the usual NCA diagrams with
respect to empty and doubly occupied local states.
This requires an infinite summation of skeleton diagrams in the
generating functional thus defining the ``Symmetrized finite-U NCA'' (SUNCA).
We show that within SUNCA the low energy scale $T_K$ (Kondo
temperature) is correctly obtained, in contrast to other simpler
approximations discussed in the literature.

\bigskip
\noindent PACS numbers: 75.20.Hr, 71.27.+a, 71.15.-m, 71.10.-w
\end{abstract}
}

\narrowtext
\section{Introduction}

Anderson impurity models
have been of considerable interest recently as generic models of local
systems with internal degrees of freedom coupled to a Fermi
gas. Although first introduced as models for magnetic impurities in
metals \cite{Hew93}, they describe two--level systems in metals \cite{Cox98},
quantum dots in mesoscopic structures \cite{Lee88,Glazman88} 
and strongly correlated lattice systems in the local approximation of 
the Dynamic Mean Field Theory (DMFT) \cite{Georges96} as well.  
In a nutshell, the Anderson model
features one or several local levels hybridizing with the conduction
electron states of the metal.  Multiple occupancy of the local levels
is inhibited by the strong Coulomb repulsion $U$ between electrons in
the local states.  As a consequence, the local levels are
approximately singly occupied, giving rise to a magnetic moment or an
equivalent degree of freedom.  Due to the coupling to the conduction
electron system, the local moment is screened \cite{Hew93}, 
or in a multi--channel situation forms a more complicated many--body 
state \cite{Andrei84}.

Most investigations of Anderson models have concentrated on the case
of infinite Coulomb repulsion $U$.  The corresponding restriction of the
local Hilbert space to electron occupation $n_d \leq 1$ allows for an
economical treatment in terms of pseudo--particle representations
\cite{Barnes76,Coleman84} and a projection onto the 
physical sector of Hilbert space.
In this framework, the simplest approximation consists of second-order
self-consistent perturbation theory in the hybridization, the
so-called Non-Crossing Approximation (NCA) \cite{Keiter71,Kuramoto83}.  
Although the
NCA has its limitations, it is a valuable tool for extracting the
complex many-body physics of Anderson impurity models.  In the single
channel case the NCA accounts correctly for the formation of a Kondo
resonance at the Fermi level below the Kondo temperature $T_K$ 
\cite{Costi96}, even
though the appearance of a local Fermi liquid state at temperatures $T
\ll T_K$ is not captured in this approximation \cite{Kroha97}.  
In the multichannel case even the correct low temperature power law
behavior is obtained in NCA \cite{CoxRuck.93}.  
However, in order to capture, e.g., the 
physics of the upper and the lower Hubbard bands in a DMFT description
of the Hubbard model and the Mott-Hubbard metal-insulator transition,
it is essential to consider the case of large but finite $U$.
It is therefore desirable to develop a
generalization of NCA to the case of finite Coulomb interaction.
In the following we present a straightforward generalization of NCA,
which conserves the symmetry of virtual transitions to the empty local
level or doubly occupied local level states. This is essential for
recovering the correct Kondo temperature $T_K$, as pointed out
by Pruschke and Grewe \cite{Pruschke89} and, as will be shown,
requires an infinite summation of a certain class of crossing diagrams.
We find that inclusion of only the first
crossing term in this resummation \cite{Pruschke89}, 
while contributing the larger part
of the change of $T_K$, is not sufficient to provide a qualitatively
correct Kondo temperature.  

\section{Pseudoparticle representation of the model}
\label{Pseudo}

The model we consider describes a local impurity level (called d-level
in the following), hybridizing with the conduction electron states.
The energy $E_d$ of the level may be located below or above the Fermi
energy. Two electrons with spins $\uparrow$ and $\downarrow$ on the
local level experience a Coulomb interaction $U$.  The local states
will be assumed to be created by pseudo--particle operators
$f_\sigma^{\dag}$ (singly occupied state with spin $\sigma$), 
$b^{\dag}$ (empty state) and $a^{\dag}$
(doubly occupied state) acting on a vacuum state without any
impurity.  We choose $f_\sigma$ to be fermion and $a$, $b$ to be boson
operators, where $b$ will be called the ``light'' and $a$ the ``heavy''
boson. 
The creation operator for the local physical electron can then
be written as    
$d^{\dag}_{\sigma}= f^{\dag}_{\sigma}b +\sigma a^{\dag} f_{-\sigma}$,
where the pseudo--particle occupation numbers must satisfy the operator
constraint
\begin{equation}
Q = a^{\dag}a + b^{\dag}b + \sum_\sigma f_\sigma^{\dag}f_\sigma = 1 \ ,
\label{2}
\end{equation}
expressing the fact that at any instant of time 
the impurity is in exactly one charge state,
empty, singly, or doubly occupied, respectively.
The fermion operators $c_{\vec k\sigma}^{\dag}$ create
electrons in conduction electron states $\mid\vec k \sigma >$ with
energy $\epsilon_k$. The Hamiltonian then takes the form 
\begin{eqnarray}
H &=& \sum_{\vec k,\sigma}\epsilon_{\vec k}c_{\vec k\sigma}^{\dag} 
c_{\vec k\sigma} + E_d (2a^{\dag}a + \Sigma_\sigma f_\sigma^{\dag}f_\sigma) 
+\nonumber
\\
&+& U a^{\dag}a + \sum_{\vec k,\sigma} V (c_{\vec k\sigma}^{\dag}
b^{\dag}f_\sigma + \sigma c_{\vec k\sigma}^{\dag}f_{-\sigma}^{\dag} a + h.c)\ ,
\label{1}
\end{eqnarray}
where $V$ is the hybridization matrix element.  
For later use we define the conduction electron density of states at the
Fermi energy as ${\cal N}(0)$ and the effective hybridization
$\Gamma = \pi {\cal N}(0) V^2$.

\section{Gauge Symmetry and Projection onto the Physical Hilbert
Space}
\label{Gauge}

The model described by the auxiliary particle Hamiltonian (\ref{1}) is
invariant under simultaneous, local $U(1)$ gauge transformations,
$f_\sigma \rightarrow f_\sigma e^{i\phi (t)}$, $b \rightarrow
be^{i\phi(t)}$, $a\rightarrow ae^{i\phi(t)}$, where $\phi(t)$ is an
arbitrary, time-dependent phase.  This gauge symmetry guarantees the
conservation of the local charge $Q$ in time.  In order to project
onto the physical subspace $Q=1$, it is therefore sufficient to carry out the
projection at time $t\rightarrow - \infty$, if the gauge symmetry is
implemented exactly. One starts with the
grandcanonical ensemble with respect to $Q$ and the associated
chemical potential $-\lambda$.  The projection is achieved by taking
the limit $\lambda \rightarrow \infty$ of any grandcanonical
expectation value of a physical operator $\hat A$ acting in the
impurity Hilbert space
\begin{equation}
\langle \hat A \rangle =  \lim_{\lambda\to\infty} \frac{\langle \hat
A \rangle _G}{\langle \hat Q \rangle _G} \ .
\label{3}
\end{equation}
Here the subscript $G$ denotes the grandcanonical ensemble.  The extra
factor $\hat Q$ in the denominator of Eq.\ (\ref{3})
has been introduced to project out the $Q = 0$
subspace. Note that in the numerator this factor can be omitted, since any
physical operator $\hat A$ acting on the impurity states consists of powers of
$d^{\dag}_{\sigma}$, $d_{\sigma}$, which annihilate any state in the 
$Q=0$ subspace, $d^{\dag}_{\sigma}|Q=0\rangle =0$, $d_{\sigma}|Q=0\rangle =0$.
A detailed description of the projection procedure is given in 
Ref.\ \cite{Costi96}.
Expectation values in the grandcanonical ensemble may be
calculated straightforwardly in perturbation theory in the
hybridization $V$, making use of Wick's theorem.  The usual
resummation techniques may be applied.  Thus the imaginary time single
particle Green's functions
\begin{equation}
G_{f\sigma} (\tau_1 -\tau_2) = - \langle \hat T [f_\sigma
(\tau_1)f_\sigma^{\dag} (\tau_2) ]\rangle_G
\label{4}
\end{equation}
and analagously for the two bosons $a,b$, may be expressed in terms of
the self--energies $\Sigma_{f,b,c}(i\omega)$ as
\begin{eqnarray}
G_{f\sigma}(i\omega) &=& [i\omega - \lambda - E_d -
\Sigma_f(i\omega)]^{-1}\nonumber \\
G_b(i\omega) &=& [i\omega - \lambda - \Sigma_b(i\omega)]^{-1}
\label{5} \\
G_a(i\omega) &=& [i\omega - \lambda - 2E_d - U - \Sigma_a(i\omega)]^{-1} \ .
\nonumber
\end{eqnarray}
The local conduction electron Green's function is given by
\begin{equation}
G_{c\sigma}(i\omega) = \Big\{\Big[G_{c\sigma}^0 (i\omega)\Big]^{-1} -\Sigma_{c\sigma}(i\omega)\Big\}^{-1}
\label{6}
\end{equation}
with 
\begin{equation}
G_{c\sigma}^0(i\omega) = \sum_{\vec k}G_{c\sigma}^0 (\vec k,i\omega)
= \sum_{\vec k} [i\omega - \epsilon_k ]^{-1} \ .
\label{7}
\end{equation}
The physical d-electron Green's function is proportional to the 
single-particle conduction electron $t$-matrix $t_{c\sigma}(i \omega )$,
and is related to the grandcanonical (unprojected) $\Sigma_{c\sigma\, G}$ as
\begin{equation}
G_{d\sigma}(i\omega) = \frac{1}{V^2}t_{c\sigma}(i\omega ) =
\frac{1}{V^2}  \lim_{\lambda\to\infty}
e^{\beta\lambda} \Sigma_{c\sigma\, G} (i\omega,\lambda)
\label{8}\ ,
\end{equation}
where $\beta$ is the inverse temperature.
The physical (projected onto the $Q=1$ subspace) local conduction
electron self--energy is then obtained from the $t$-matrix as 
\begin{equation}
\Sigma_{c\sigma}(i\omega) = \frac{V^2G_{d\sigma}(i\omega)}
{1+V^2G^0_{c\sigma}(i\omega)G_{d\sigma}(i\omega)}
\label{9}\ .
\end{equation}

\section{Generating Functional}
\label{generating functional}

Gauge invariant approximations conserving the local charge $Q$ may be
derived from a Luttinger--Ward generating functional $\Phi$.  
For a given approximation
the functional $\Phi$ is defined by a sum of closed skeleton
diagrams. The self--energies $\Sigma_\mu, \mu = a,b,f,c$, are
obtained by taking the functional derivatives
\begin{equation}
\Sigma_{\mu} = \frac{\delta \Phi}{\delta G_\mu} \ .
\label{10}
\end{equation}
The ``Non-crossing Approximation'' (NCA) in the limit $U\rightarrow
\infty$ is defined by the single lowest order diagram (2nd order in
$V$) containing a light boson line (the first diagram of
Fig. 1). In the limit of small hybridization element $V_{\vec k}$,
it appears to be justified to keep only the lowest order contribution
in $V$.  However, as discussed in Refs.\ \cite{Kroha97,Kroha01}, 
the singular behavior
of vertex functions may require to include these as well. This turns
out to be necessary in the single channel model where the formation of
a many-body resonance state is essential for recovering the Fermi
liquid behavior, and less so in the multi-channel models.
Including an infinite class of skeleton diagrams in $\Phi$ (in a ``Conserving
T-matrix Approximation'' (CTMA)), which allows to capture a singular
structure in the spin and charge excitation sectors, the low temperature Fermi
liquid phase of the single channel Anderson model is recovered 
\cite{Kroha97}.

Here we are interested in constructing a simpler generalization of NCA
to describe the case of finite $U$. It seems straightforward
to define such an approximation on the NCA level by adding to the second order
skeleton diagram for $\Phi$ containing the light boson
(the first diagram in Fig. 1 a)) 
the corresponding diagram containing the heavy boson (the second
diagram in Fig. 1 a)).  This approximation and certain extensions motivated by 
perturbative arguments \cite{Pruschke89} or by a $1/N$ expansion 
($N$ beeing the spin degeneracy) \cite{Schoenhammer89}
have been considered sometime ago. 
However, in the case of finite $U$ NCA was found to fail badly: 
Not even the Kondo energy scale is
recovered in the so-defined approximation.  The reason for this
failure is obvious: 
In the Kondo regime $(n_d {\buildrel \sim \over <}
1)$ the local spin is coupled to the conduction electron spin density
at the impurity through the antiferromagnetic exchange coupling
\begin{equation}
J = V^2 (- \frac{1}{E_d} + \frac{1}{E_d+U}).
\label{11}
\end{equation}
 The two terms on the r.h.s. of this relation arise from virtual
 transitions into the empty and doubly occupied local level, which e.g.
 contribute equally in the symmetric case $\mid E_d \mid = E_d + U$.
 The symmetric occurrence of these two virtual processes in all
 intermediate states is not included in the simple extension of NCA
 proposed above. Rather, the self--energy insertions in each of the two
diagrams contain always only one of the processes, leading to an
effective $J$ which is only one half of the correct value.
Correspondingly, the Kondo temperature $T_K \sim \exp [-1/(2{\cal
N}(0)J)]$ comes out to scale as the square of the correct value, which
can be orders of magnitude too small.  

\begin{figure}
\vspace*{-0cm}
\centerline{\psfig{figure=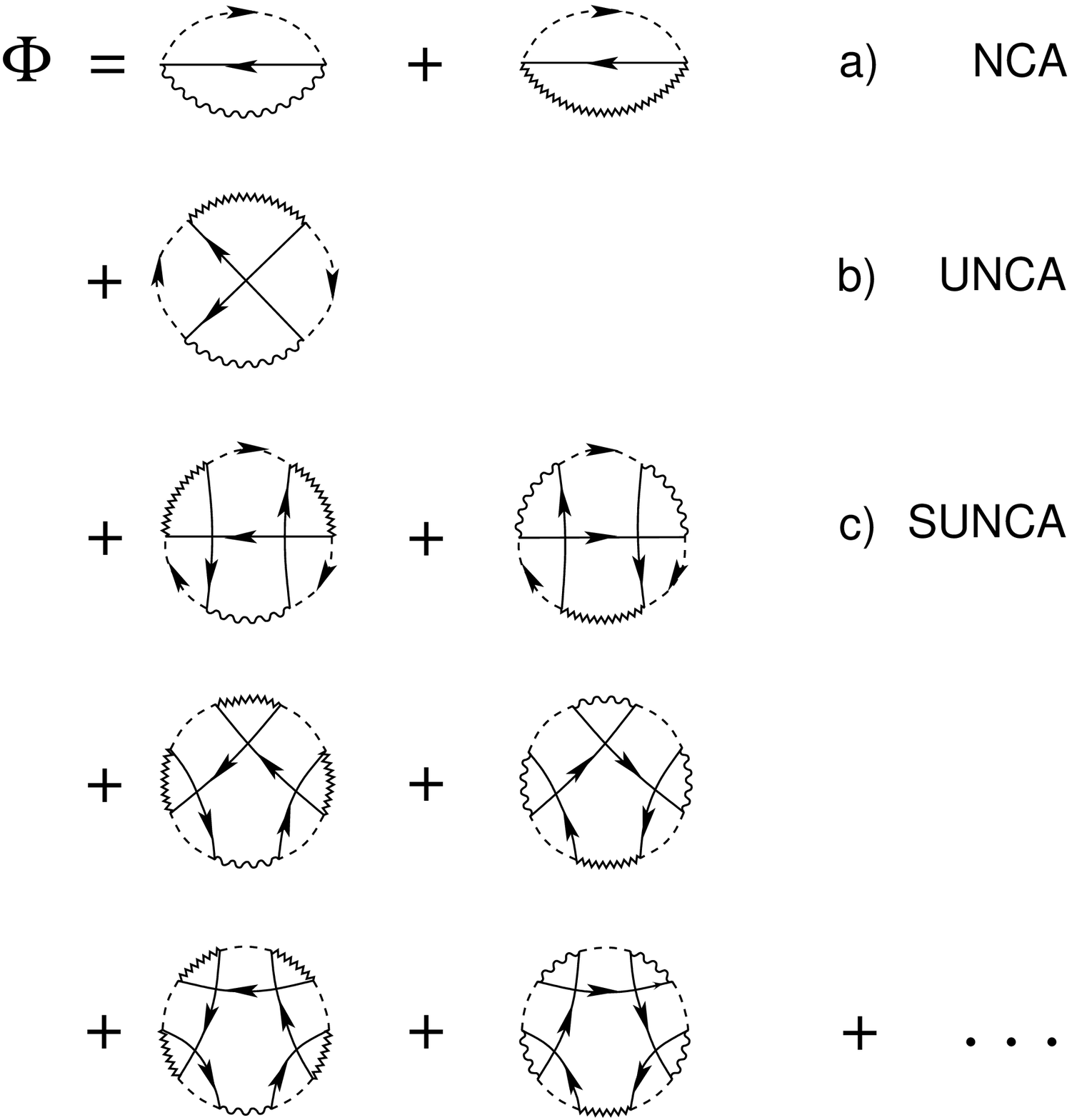,width=0.98\linewidth}}
\vspace*{0.5cm}
\caption{
Diagrammatic representation of the generating functional to describe
the Anderson impurity model at finite $U$. Throughout this paper,
solid, dashed, wiggly and zig-zag
lines correspond to conduction electron $c$, pseudo--fermion $f$, 
light boson $b$, and heavy boson $a$ propagators, respectively.
a) NCA including light and heavy boson lines.
a)--b) Finite-$U$ NCA (UNCA). This approximation amounts to 
renormalizing only one of the vertices in each of the
self--energy diagrams of Fig.\ \ref{selfenergy} and keeping only one
(light or heavy boson) rung in the corresponding vertex function 
(see text). a)--c) Symmetrized finite-$U$ NCA (SUNCA).  
\label{genfunct}}
\end{figure}

To correct this deficiency it is necessary to include additional 
diagrams, restoring the symmetry between the two virtual processes.
As a first step one may add the next order skeleton diagram to $\Phi$
(Fig.\ \ref{genfunct} b)). 
As we will show below, this approximation,
later referred to as UNCA, helps to recover a large part of the
correct behavior of $T_K$  \cite{Pruschke89}. 
However, as seen from the preceding discussion,
for a completely symmetric treatment of empty and doubly occupied 
intermediate
states one must first consider the diagrams of bare perturbation theory
instead of skeleton diagrams.  
A symmetric class of diagrams is generated by replacing a light boson line 
with a heavy boson line in each of the bare (non-skeleton) 
diagrams comprising the NCA, and vice versa.   
Each replacement leads to a crossing of conduction electron lines spanning 
one fermion and at most two boson lines. 
A conserving approximation is then constructed by substituting 
renormalized propagators for the bare ones 
and keeping only skeleton diagrams. 
The resulting generating functional $\Phi$ is shown diagrammatically 
in Fig.\ \ref{genfunct} a)--c). 
\noindent
These diagrams look similar to the CTMA
diagrams mentioned above, but contain one light boson line and an
arbitrary number of heavy boson lines, or vice versa.  Diagrams with,
for example, two light boson lines and an arbitrary number of heavy
boson lines (and conduction electron lines spanning at most one
fermion line) are reducible and do not appear. We will call the
approximation defined by the generating functional given by the sum of
the diagrams of Fig \ref{genfunct} 
``Symmetrized finite-U NCA''
(SUNCA). The above
approximation corresponding to the CTMA at $U\rightarrow\infty$, termed
``Symmetrized finite-U Conserving T-matrix Approximation'' (SUCTMA) 
is thus defined in a natural way by summing up all skeleton $\Phi$ diagrams  
containing a single closed ring of auxiliary particle propagators
with an arbitrary number of light or heavy boson lines, 
dressed by conduction electron
lines spanning only one fermion line.  
Thus, the SUCTMA is defined by adding to the
diagrams of the SUNCA the CTMA diagrams  with (only) light
boson lines or (only) heavy boson lines.  The SUCTMA equations have not
yet been evaluated.

\section{Results of SUNCA}
\label{results}

\begin{figure}
\vspace*{-0cm}
\centerline{\psfig{figure=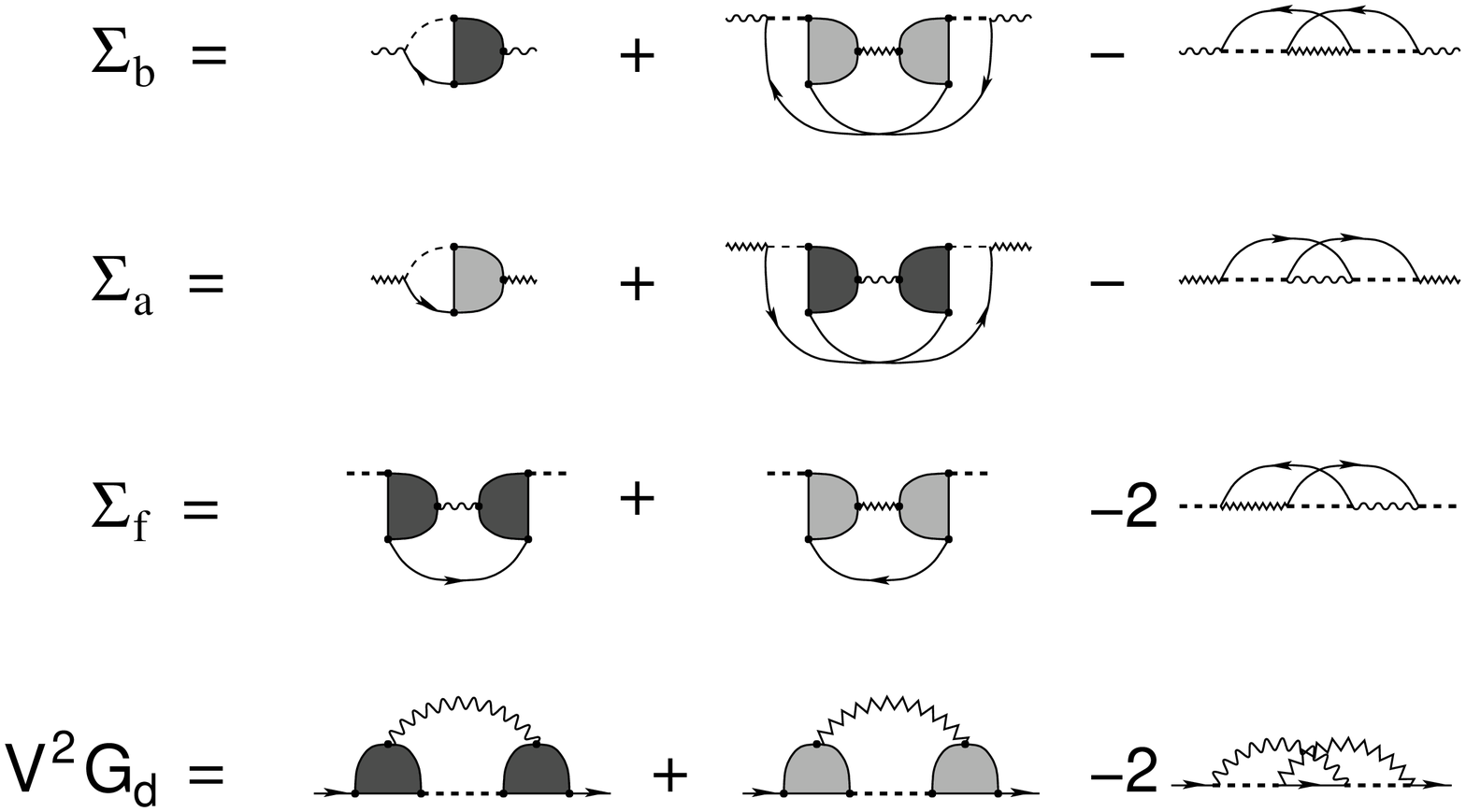,width=1.0\linewidth}}
\vspace*{0.5cm}
\caption{
Diagrammatic representation of the auxiliary particle self--energies
of SUNCA in terms of the renormalized hybridization vertices, defined
in Fig.\ \ref{vertex}. In each line the third diagram is subtracted
in order to avoid double counting of terms within the first two 
diagrams.
\label{selfenergy}}
\end{figure}

\begin{figure}
\vspace*{-0cm}
\centerline{\psfig{figure=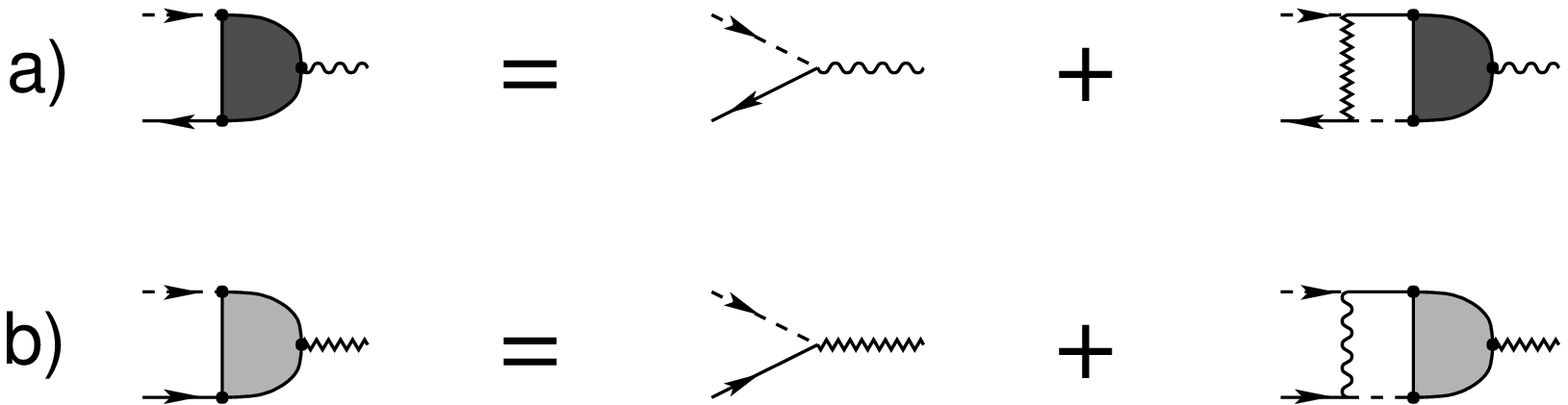,width=1.0\linewidth}}
\vspace*{0.5cm}
\caption{
Diagrammatic representation of the Bethe--Salpeter equations for a) the
renormalized light boson (empty impurity) and b) the
renormalized heavy boson (doubly occupied impurity) vertex, as
generated by the SUNCA Luttinger--Ward functional (Fig.~1).
\label{vertex}}
\end{figure}

As discussed above, the self--energies $\Sigma_\mu$ are obtained by
functional differentiation of the generating functional with respect to
the Green's functions $G_\mu$. The functional $\Phi$ defined by
Fig.\ \ref{genfunct} leads to an infinite series of diagrams 
for $\Sigma_\alpha$, which may be conveniently presented in terms of 
three--point vertex functions (the
filled semicircles with three legs: one boson and two fermion lines),
see Fig.\ \ref{selfenergy}. It is necessary to substract a diagram 
of 4th order in $V$ in each case to avoid double counting. 
On the level of SUNCA and SUCTMA the vertex functions consist of ladder 
summations, with light or heavy boson lines as rungs, and are defined
diagrammatically in Fig.\ \ref{vertex}. Note that keeping only a single 
light or heavy boson rung in these vertex functions corresponds to UNCA.

The expressions for the self--energies $\Sigma_\mu$ defined by 
Figs.\ \ref{selfenergy} and \ref{vertex}, 
together with the definition of the Green's functions, 
Eqs. (\ref{5}),
constitute a set of nonlinear integral equations for $\Sigma_\mu
(\omega), \mu = a,b,f$.  The local conduction electron self--energy
$\Sigma_c$ does not appear in any internal Green's functions
because it contains at least one auxiliary particle loop, i.e.
carries a factor ${\rm exp}(-\beta \lambda)$ and thus 
\noindent
vanishes 
due to the exact projection onto the physical Hilbert space 
($\lambda \to \infty$) \cite{Costi96}.
$G_d$ and therefore $\Sigma_c$ may be calculated at the end
via Eqs.\ (\ref{8}), (\ref{9}) by using
the self--consistently determined $G_\alpha, \alpha = a,b,f$.  
The SUNCA equations are given explicitly in appendix A.
Although 
these equations are more involved than the regular NCA,
they are numerically considerably more easily tractable than the 
CTMA equations \cite{Kroha97,Kroha01}, 
since SUNCA contains only renormalized three-point
vertices (see Fig.\ \ref{selfenergy}) as compared to four-point vertex 
functions occuring in CTMA \cite{Kroha97}.  
We solved the SUNCA equations numerically in the real frequency 
representation, i.e. after analytic continuation from Matsubara 
frequencies $\omega_n$ to the real frequency axis. 

As a first important characteristic feature of the pseudo--particles we
note that the single--particle excitation spectrum is powerlaw
divergent, $G_\mu(\omega) \sim \omega^{-\alpha_\mu}, \mu = a,b,f$
reflecting the abundance of low energy excitations forced by the
constraint. At finite temperature $T$ these singularities are cut off
at the scale $\omega \sim T$. As observed in earlier work 
\cite{Kroha97,Costi94},
the values of the exponents $\alpha_\mu$ are characteristic of the
state of the system.  In the single channel case, when the ground
state is a local Fermi liquid, the exponents may be inferred from the
Friedel sum 
\begin{figure}
\vspace*{-0cm}
\centerline{\psfig{figure=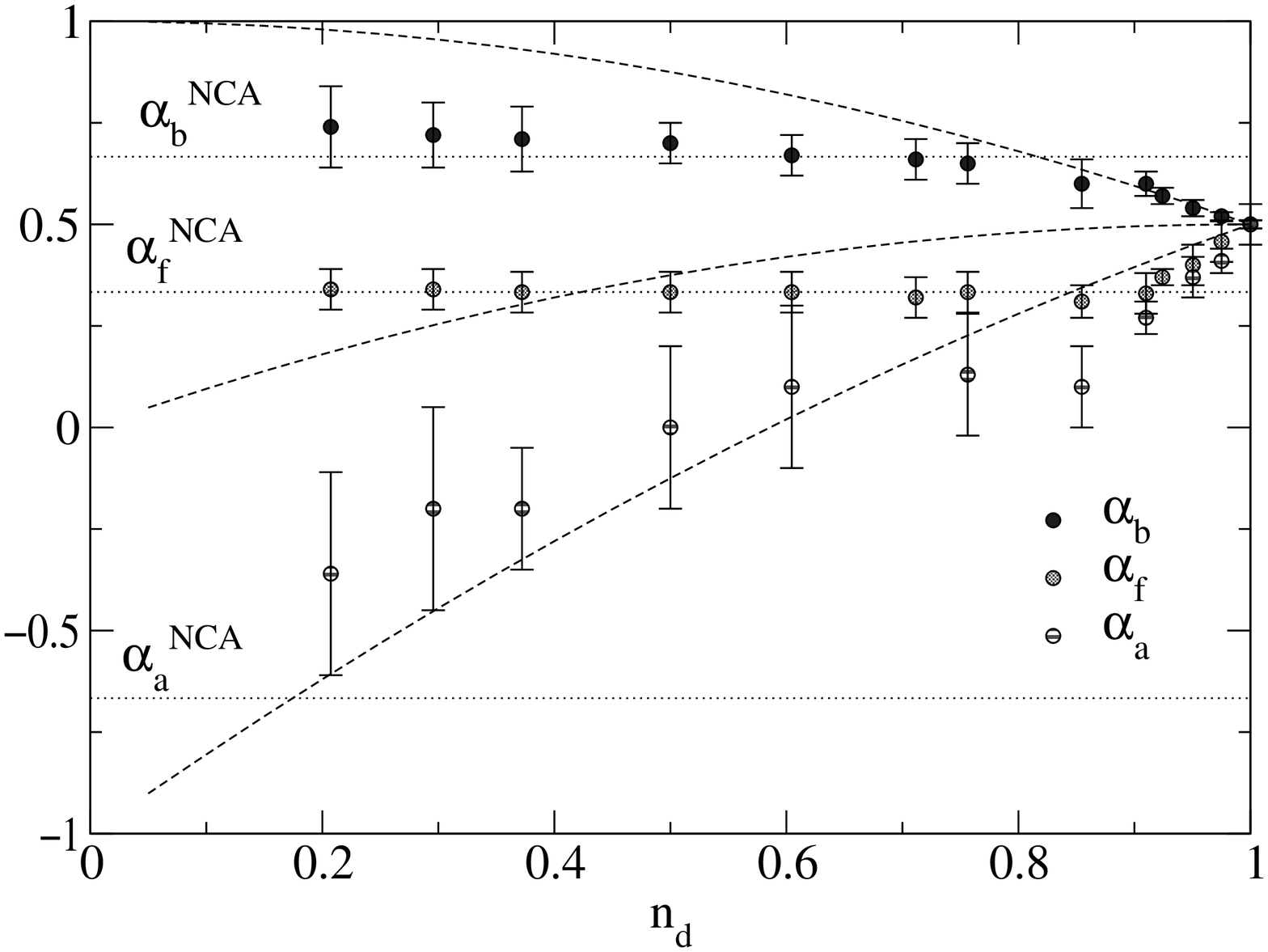,width=0.95\linewidth}}
\vspace*{0.5cm}
\caption{
Infrared threshold exponents of the auxiliary particle 
spectral functions $-{\rm Im} G^R_{f,a,b}(\omega )/\pi$, 
in dependence of the impurity occupation $n_d$, for fixed values of
$\Gamma$ and $U$ and varying $E_d$.
Dashed curved lines: exact results (Eq.~(12)); horizontal lines:
NCA results; data points with error bars: SUNCA results.
In the Kondo limit ($n_d \to 1$)
the exact exponents are recovered, while in the mixed valence and
empty impurity regime the SUNCA results for $\alpha _f$ and $\alpha _b$
cross over to the NCA values.
\label{exponents}}
\end{figure}\noindent
rule relating the scattering phase shifts
$\eta_{\mu,\sigma}$ to the number of electrons bound to the impurity
in each channel $\Delta n_{\mu,\sigma}=\eta_{\mu\sigma}/\pi$. The
exponents $\alpha_\mu$ in turn
are related to the $\eta_{\mu,\sigma}$ by the general result first
derived for the x-ray edge singularities \cite{Nozieres69}, 
$\alpha_\mu = 1 - \sum_\sigma(\eta_{\mu\sigma}/\pi)^2$.
This is so, since e.g. the heavy boson Green's function, describing the
transition amplitude for a doubly occupied impurity to be created at time
$t=0$ and removed at a later time is proportional to
the overlap of the free Fermi sea of conduction electrons with the
ground state of the Anderson model.  The change in the occupation of
the local level due to the hybridization with the conduction band 
$\Delta n_{\mu\sigma}$ 
is the difference between the $t=0$ initial
impurity occupation $n _{\mu \sigma} (t=0)$ (without hybridization)
and the occupation in the ground state of the Anderson impurity 
model $n_{d\sigma}=n_d/2$, i.e. it depends on the initial conditions of the
different Green's functions $G_{\mu}$, $\mu=a,b,f$. 
Thus we have $\Delta n_{a,\sigma} = \frac{2-n_d}{2}, \Delta
n_{f,\sigma} = \delta_{\sigma,\sigma_0}-\frac{1}{2}n_d$ (where
$\sigma_0$ is the spin of the fermion in the Green's function $G_{f\sigma_0}$), 
and $\Delta n_{b,\sigma}= - \frac{n_d}{2}$.
The infrared threshold exponents of $G_\mu(\omega)$ are
therefore given by
\begin{eqnarray}
\alpha_a  &=&  - 1 + 2n_d - \frac{n_d^2}{2} \nonumber \\
\alpha_b &=& 1 - \frac{n_d^2}{2}\label{12} \\
\alpha_f &=& n_d - \frac{n_d^2}{2} \ . \nonumber
\end{eqnarray}

In Fig.\ \ref{exponents} 
we show the exponents $\alpha_\mu$ for different $n_d$, as
obtained from a numerical solution of the SUNCA equations.  Also
shown is the exact result given by Eq.\ (\ref{12}) (dashed lines), and the
analytical result that can be extracted analytically from the NCA equations
(defined by the first two diagrams in Fig. 1) in analogy to 
Ref.\ \cite{Muha88}.  The
numerical results of SUNCA (data points) are seen to approach the
exact result in the limit $n_d \rightarrow 1$, but in the case of
$\alpha_b$ and $\alpha_f$ appear to follow the NCA result rather than
the exact result for $n_d \leq 0.8$.  The results 
\begin{figure}
\vspace*{-0cm}
\centerline{\psfig{figure=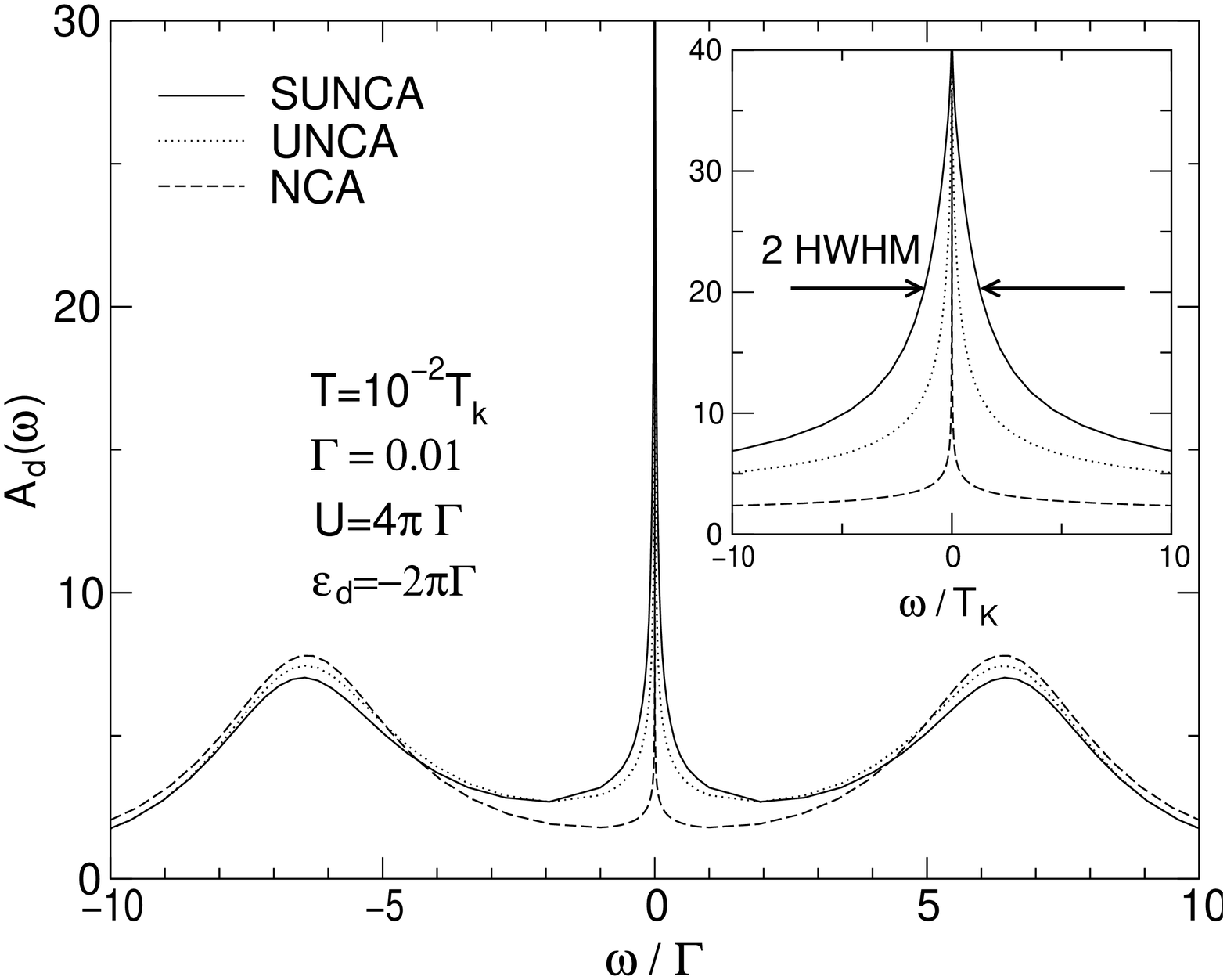,width=1.0\linewidth}}
\vspace*{0.5cm}
\caption{
Local electron spectral function calculated using
NCA, UNCA, and SUNCA. The Kondo temperature is determined as the 
HWHM of the Kondo peak (see inset). 
It is seen that in NCA the Kondo peak width 
comes out orders of magnitude too low.
\label{spectrum}}
\end{figure} \noindent
for the exponent
$\alpha_a$ trace the exact behavior in reasonable agreement.
Clearly the SUNCA does much better than the simple NCA. From our
experience \cite{Kroha97} with the Anderson model in the 
limit $U\rightarrow \infty$,
we expect that the correct exponents should be recovered in SUCTMA.

We now turn to the d-electron spectral function $A_d(\omega) =
\frac{1}{\pi} {\rm Im} G_d(\omega - i0)$.  
Fig.\ \ref{spectrum} shows the results for
$A_d(\omega)$ for the symmetric Anderson model in the Kondo regime
($n_d \approx 1$) at a very low temperature of $T\simeq
10^{-2}T_K$.  Shown are the results obtained from the simple NCA
(diagrams of first line in Fig. 1), the perturbatively corrected
version UNCA (including the diagram in the second line of Fig. 1) and
the full SUNCA.  The inset shows that the width of the Kondo resonance
peak, which is a measure of $T_K$, comes out orders of magnitude
different in the three approximations.  In order to compare the
numerical results with the exact expression for $T_K$, 
\begin{equation}
T_K = \rm{min} \Big\{\frac{1}{2\pi}U \sqrt{I}, \sqrt{D\Gamma}\Big\}e^{-\pi/I}
\label{13}
\end{equation}
where
\begin{equation}
I=2\left[\frac{\Gamma}{|E_d|}+\frac{\Gamma}{E_d + U}\right] \ ,
\label{14}\
\end{equation}
we determine $T_K$ as the half width of the Kondo resonance
at half maximum (HWHM).
\begin{figure}
\vspace*{-0cm}
\centerline{\psfig{figure=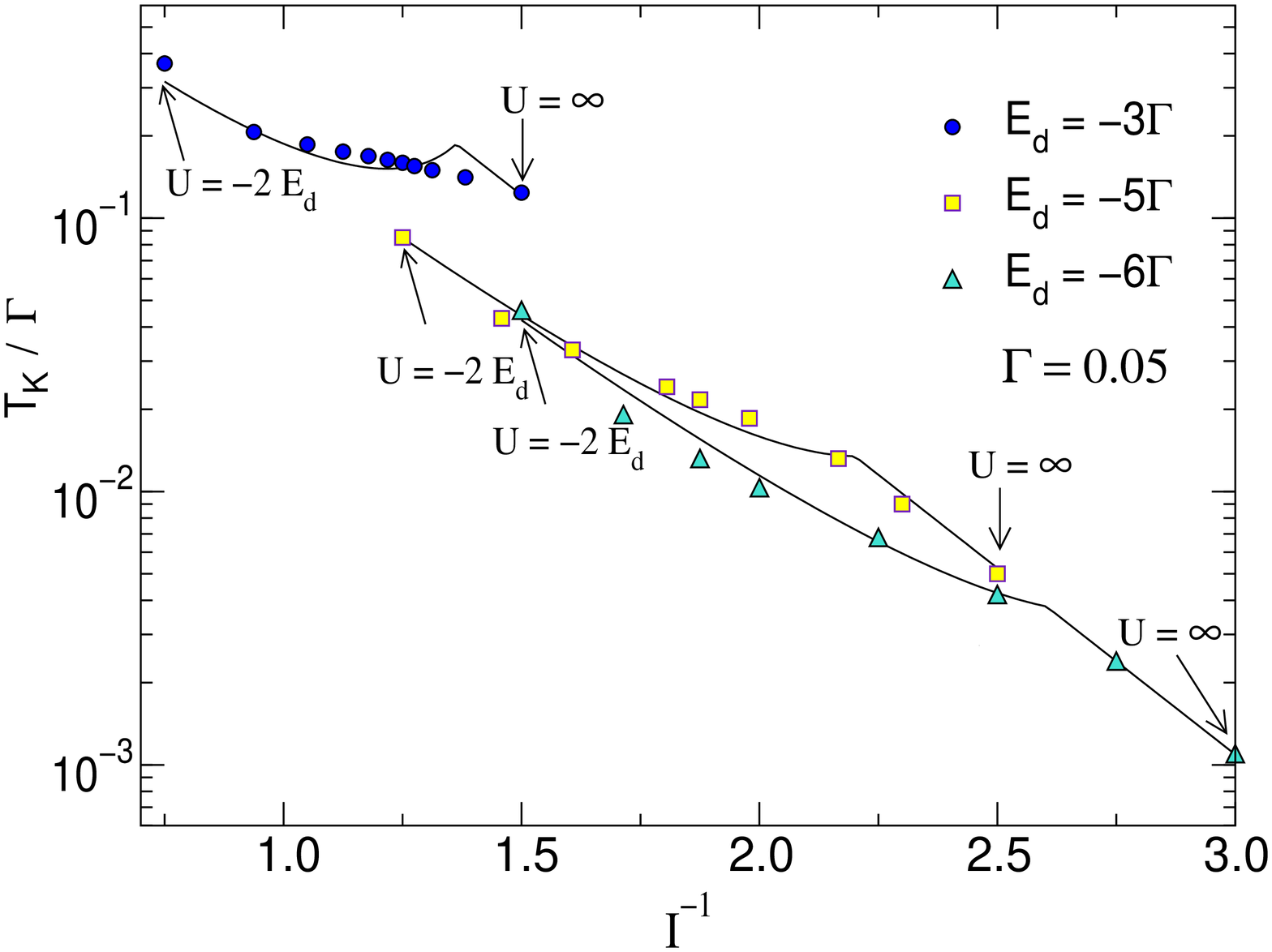,width=1.0\linewidth}}
\vspace*{0.5cm}
\caption{
Kondo Temperature for various parameters
$E_d$, $U$ and fixed $\Gamma$. Solid lines represent the exact results, 
Eqs.\ (13), (14). Data points are the SUNCA results
determined from the width of the Kondo peak in the d-electron
spectral function.   
\label{TKU}}
\end{figure}

In Fig.\ \ref{TKU} 
the results for $T_K/\Gamma$ obtained in this way for a
fixed value of $\Gamma = 0.05$ (in units of $D$) and several values of
$E_d/\Gamma$, as a function of $I(U/\Gamma)$ 
(data points) are compared with the exact values
Eqs.\ (\ref{13}), (\ref{13}) (solid lines).  The agreement is excellent,
demonstrating that the SUNCA provides the correct scale $T_K$ for a
wide range of parameters $E_d$ and $U$.
\section{Conclusion}
\label{conclusion}

In this paper we have proposed a conserving scheme to describe the
Anderson impurity model at finite on-site repulsion $U$ within the auxiliary
particle method. In order to incorporate the correct value of the spin
exchange coupling $J$ into the theory and, hence, to obtain the correct
size of the low-energy scale $T_K$, it is necessary to treat fluctuation
processes into the empty and into the doubly occupied intermediate 
state on equal footing at the level of bare perturbation theory. 
The simplest Luttinger--Ward functional which is completely symmetric in
this respect consists of an infinite series of skeleton diagrams,
corresponding to ladder--type vertex renormalizations in the
self--energies. Although considerably more involved than the regular NCA,
this approximation, termed ``symmetrized finite--$U$ NCA'' (SUNCA), 
is numerically tractable on a typical workstation.
We find that SUNCA recovers the correct 
Kondo temperature over a wide range of the parameters of the Anderson
model $E_d$, $U$, and $\Gamma$, while simplified approximations (NCA, UNCA)
produce a low-energy scale typically orders of magnitude smaller than the
exact value. This result is especially relevant for a correct description 
of the low temperature properties of 
strongly correlated lattice models by a diagrammatic many-body 
technique, since in the limit of infinite dimensions these models
reduce to a selfconsistent, finite-$U$ single-impurity problem. 
Applications of the present theory to such models are currently
in progress.

We are grateful to T. A. Costi and Th. Pruschke for useful discussions.
This work was supported by DFG through Sonderforschungsbereich 195 
(S.K., J.K., P.W.) and by the ESF program ``Fermi liquid instabilities 
in correlated metals'' (FERLIN).

\end{multicols}

\vspace*{1cm}

\begin{appendix}
\narrowtext

\section{SUNCA equations}

In this appendix we explicitly give the self-consistent SUNCA equations
which, together with the definitions Eqs.\ (\ref{5}) of the Green's 
functions, determine the auxiliary particle self-energies. 
We also give the expression for the physical d-electron spectral function
in terms of the auxiliary particle propagators.

We first define the ladder vertex functions $T_a$, $T_b$ with 
heavy boson $a$ and light boson $b$ rungs, respectively, as
shown diagrammatically in Fig.\ \ref{vertex}. These vertex functions,
projected onto the physical subspace $Q=1$ and analytically continued 
to real frequencies, obey the following Bethe--Salpeter equations,
\widetext
\top{-2.8cm}
\begin{eqnarray}
T_{a\sigma}(\omega,\Omega) &=& \Gamma \int{{d\epsilon\over\pi}
f(\epsilon-\Omega) A^0_{c-\sigma}(\epsilon-\Omega) G_{f-\sigma}(\epsilon)
G_a(\epsilon+\omega-\Omega)} \nonumber \\
&+&\Gamma \int{{d\epsilon\over\pi}
f(\epsilon-\Omega) A^0_{c-\sigma}(\epsilon-\Omega) G_{f-\sigma}(\epsilon)
G_a(\epsilon+\omega-\Omega) T_{a-\sigma}(\epsilon,\Omega)}\\
T_{b\sigma}(\omega,\Omega) &=& \Gamma \int{{d\epsilon\over\pi}
f(\epsilon-\Omega) A^0_{c\sigma}(\Omega-\epsilon) G_{f-\sigma}(\epsilon)
G_b(\epsilon+\omega-\Omega)} \nonumber\\
&+&\Gamma \int{{d\epsilon\over\pi}
f(\epsilon-\Omega) A^0_{c\sigma}(\Omega-\epsilon) G_{f-\sigma}(\epsilon)
G_b(\epsilon+\omega-\Omega) T_{b-\sigma}(\epsilon,\Omega)} \ ,
\end{eqnarray}
\bottom{-2.7cm}
\narrowtext
where $f(\epsilon)$ is the Fermi function,
$A^0_{c\sigma}(\epsilon)={1 \over \pi} {\rm Im} G_{c\sigma}^0(\epsilon)
/{\cal N}(0)$
the bare conduction electron density of states per spin, normalized to the
density of states at the Fermi level and, for concreteness,
all propagators are to be understood as the advanced ones.
The auxiliary particle self--energies (Fig.\ \ref{selfenergy}) 
are then given by,
\widetext
\top{-2.8cm}
\begin{eqnarray}
\Sigma_{f\sigma}(\omega) &=& \Gamma \int{{d\epsilon\over\pi}
  f(\epsilon-\omega) A^0_{c\sigma}(\omega-\epsilon) G_b(\epsilon)
  [1+T_{a\sigma}(\omega,\epsilon)]^2} +
  \Gamma \int{{d\epsilon\over\pi}
  f(\epsilon-\omega) A^0_{c-\sigma}(\epsilon -\omega) G_a(\epsilon)
  [1+T_{b\sigma}(\omega,\epsilon)]^2}  \nonumber\\
&-&2\Gamma ^2 \int{{d\epsilon\over\pi} f(\epsilon-\omega) 
  A^0_{c\sigma}(\omega-\epsilon) G_b(\epsilon)
  \int{{d\epsilon^\prime\over\pi} f(\epsilon^\prime-\epsilon)
  A^0_{c-\sigma}(\epsilon^\prime-\epsilon)
  G_{f-\sigma}(\epsilon^\prime)G_a(\epsilon^\prime+\omega-\epsilon)}}
\end{eqnarray}
\begin{eqnarray}
\Sigma_b(\omega) &=& \Gamma  \sum_{\sigma} \int{{d\epsilon\over\pi}
  f(\epsilon-\omega) A^0_{c\sigma}(\epsilon-\omega) G_{f\sigma}(\epsilon)
  [1+T_{a\sigma}(\epsilon,\omega)]} \nonumber \\
  &+&  \Gamma ^2 \sum_{\sigma}  \int{{d\epsilon\over\pi}
  f(\epsilon-\omega) A^0_{c\sigma}(\epsilon-\omega) 
  G_{f\sigma}(\epsilon)} \times \\
&&  \int{{d\epsilon^\prime\over\pi} 
  f(\epsilon^\prime-\omega) A^0_{c-\sigma}(\epsilon^\prime-\omega)
  G_{f-\sigma}(\epsilon^\prime)G_a(\epsilon^\prime+\epsilon-\omega)
  \left\{[1+T_{b\sigma}(\epsilon,\epsilon^\prime+\epsilon-\omega)]\;
  [1+T_{b-\sigma}(\epsilon^\prime,\epsilon^\prime+\epsilon-\omega)]-1\right\}}
  \nonumber
\end{eqnarray}
\begin{eqnarray}
\Sigma_a(\omega) &=& \Gamma \sum_{\sigma} \int{{d\epsilon\over\pi}
  f(\epsilon-\omega) A^0_{c-\sigma}(\omega-\epsilon) G_{f\sigma}(\epsilon)
  [1+T_{b\sigma}(\epsilon,\omega)]} \nonumber \\
  &+&  \Gamma^2 \sum_{\sigma} \int{{d\epsilon\over\pi}
  f(\epsilon-\omega) A^0_{c-\sigma}(\omega-\epsilon) 
  G_{f\sigma}(\epsilon)}\times \\
&& \int{{d\epsilon^\prime\over\pi} 
  f(\epsilon^\prime-\omega) A^0_{c\sigma}(\omega-\epsilon^\prime)
  G_{f-\sigma}(\epsilon^\prime)G_b(\epsilon^\prime+\epsilon-\omega)
  \left\{[1+T_{a\sigma}(\epsilon,\epsilon^\prime+\epsilon-\omega)]\;
  [1+T_{a-\sigma}(\epsilon^\prime,\epsilon^\prime+\epsilon-\omega)]-1\right\}}\ .
  \nonumber
\end{eqnarray}

In order to calculate the physical impurity electron spectral function 
$A_{d\sigma}$ from the selfconsistently determined $G_a$, $G_b$, $G_f$, 
it is convenient to define modified vertex functions as 
\begin{eqnarray}
S^R_{a\sigma}(\omega,\Omega) &=& 1 + \Gamma  \int{{d\epsilon\over\pi}
  f(\epsilon-\Omega) A^0_{c\sigma}(\epsilon-\Omega)\;  {\rm Re} \{
  G_{f\sigma} (\epsilon)[1+T_{a\sigma}(\epsilon,\Omega)]\}\; 
  G_a(\epsilon+\omega)}\\
S^I_{a\sigma}(\omega,\Omega) &=& 1 + \Gamma  \int{{d\epsilon\over\pi}
  f(\epsilon-\Omega) A^0_{c\sigma}(\epsilon-\Omega)\;  {\rm Im} \{
  G_{f\sigma} (\epsilon)[1+T_{a\sigma}(\epsilon,\Omega)]\}\; 
  G_a(\epsilon+\omega)}\\
S^R_{b\sigma}(\omega,\Omega) &=& 1 +\Gamma  \int{{d\epsilon\over\pi}
  f(\epsilon-\Omega) A^0_{c-\sigma}(\Omega-\epsilon) \; {\rm Re}\{ 
  G_{f\sigma}(\epsilon)[1+T_{b\sigma}(\epsilon,\Omega)]\}\; 
  G_b(\epsilon-\omega)}\\
S^I_{b\sigma}(\omega,\Omega) &=& 1 +\Gamma  \int{{d\epsilon\over\pi}
  f(\epsilon-\Omega) A^0_{c-\sigma}(\Omega-\epsilon) \; {\rm Im}\{
  G_{f\sigma}(\epsilon)[1+T_{b\sigma}(\epsilon,\Omega)]\}\; 
  G_b(\epsilon-\omega)} \ .
\end{eqnarray}
The impurity spectral function then reads
\begin{eqnarray}
A_{d\sigma}(\omega) &=& -{1\over\pi} {\rm Im}
    {\int \frac{d\Omega}{\pi} \; \frac{{\rm e}^{-\beta\Omega}}{f(-\omega)}\; 
    G_{f\sigma}(\Omega+\omega)\left\{{\rm Im} [G_b(\Omega)]
    [S^R_{a-\sigma}(\omega,\Omega)^2-S^I_{a-\sigma}(\omega,\Omega)^2]+ 
    2 {\rm Re} [G_b(\Omega)]
    S^R_{a-\sigma}(\omega,\Omega) S^I_{a-\sigma}(\omega,\Omega) \right\} }   
    \nonumber\\
&& -{1\over\pi}  {\rm Im}{\int \frac{d\Omega}{\pi}\;  
    \frac{{\rm e}^{-\beta\Omega}} {f(\omega)}\; 
    G_{f-\sigma}(\Omega-\omega) \left\{ {\rm Im} [G_a(\Omega)]
    [S^R_{b\sigma}(\omega,\Omega)^2-S^I_{b\sigma}(\omega,\Omega)^2]+ 
    2 {\rm Re} [G_a(\Omega)]
    S^R_{b\sigma}(\omega,\Omega) S^I_{b\sigma}(\omega,\Omega) \right\}} 
    \nonumber \\
&& +2{\Gamma\over\pi} \int{\frac{d\Omega}{\pi} \; 
    \frac{{\rm e}^{-\beta\Omega}}{f(\omega)}\;  \int{
    {d\epsilon\over \pi} f(\epsilon-\Omega)
    A^0_{c-\sigma}(\epsilon-\Omega) \; 
    {\rm Im} [G_b(\Omega) G_{f-\sigma}(\epsilon)]\;  
    {\rm Im} [G_{f\sigma}(\Omega+\omega) G_a(\epsilon+\omega)] }} \ .
\label{Ad}
\end{eqnarray}
\narrowtext
Note that the exponential divergencies of the  statistical factors appearing 
in Eq.\ (\ref{Ad}) are compensated by the threshold behavior
of the corresponding auxiliary particle spectral functions 
$A_{\mu}(\omega ) = \frac{1}{\pi} {\rm Im} G_{\mu}(\omega)$, $\mu =a,b,f$ 
in the integrands. 
For the numerical treatment, these divergencies can be 
explicitly absorbed by formulating the self--consistency equations
(A1)--(A10) in terms of the functions $\tilde A_{\mu}(\omega)$ which are
defined via 
\begin{equation}
A_{\mu}(\omega ) = f(-\omega) \tilde A_{\mu}(\omega) 
\end{equation}  
and, hence, have no exponential divergence.
We thus have, e.g., $\exp (-\beta\omega) A_{\mu}(\omega ) = f (\omega ) 
\tilde A_{\mu}(\omega)$. Details of this representation are described in
Ref.\ \cite{Costi96}.

\end{multicols}

\end{appendix}

\narrowtext

\end{multicols}
\end{document}